\documentclass[useAMS,usenatbib,usegraphicx]{PoS}
\usepackage[authoryear,square]{natbib}
\bibpunct{(}{)}{;}{a}{}{,}

\usepackage{multirow}
\title{Cosmology from HI galaxy surveys with the SKA}

\vspace*{-10.0mm}

\ShortTitle{Cosmology from galaxy surveys with the SKA}

\author{\speaker{Filipe B. Abdalla}$^{1,2}$, Philip Bull$^3$, Stefano Camera$^4$, Aur{\'e}lien Benoit-L{\'e}vy$^{1}$, Benjamin Joachimi$^1$, Donnacha Kirk$^1$, Hans-Rainer Kl\"ockner$^5$,  Roy Maartens $^{6,7}$, Alvise Raccanelli$^{8,9}$, Mario G. Santos$^{4,6,10}$, Gong-Bo Zhao$^{11,7}$ on behalf of the Cosmology SWG.\\

$^1$Department of Physics and Astronomy, University College London,
London WC1E
6BT, UK\\
$^2$Department of Physics and Electronics, Rhodes University, PO Box 94, Grahamstown, 6140 South Africa; 
$^3$ Institue of Theoretical Astrophysics, University of Oslo, P.O. Box 1029
Blindern, N-0315 Oslo, Norway; $^4$ CENTRA, Instituto Superior T\'{e}cnico, Universidade de Lisboa, Lisboa 1049-001, Portugal; $^5$ Max-Planck-Institut fur Radioastronomie, Auf dem H\"ugel 69, 53121 Bonn, Germany; $^6$ Department of Physics, University of Western Cape, Cape Town 7535, South Africa; $^7$ Institute of Cosmology \& Gravitation, University of Portsmouth, Portsmouth PO1 3FX, UK; $^8$ Jet Propulsion Laboratory, California Institute of Technology, Pasadena CA 91109, USA; $^9$ California Institute of Technology, Pasadena CA 91125, USA; $^{10}$ SKA SA, 4rd Floor, The Park, Park Road, Pinelands, 7405, South Africa; $^{11}$ National Astronomy Observatories, Chinese Academy of Sciences, Beijing, 100012, P.R.China

E-mail:\email{fba@star.ucl.ac.uk}

}
\abstract{The Square Kilometer Array (SKA) has the potential to produce galaxy redshift surveys which will be competitive with other state of the art cosmological experiments in the next decade. In this chapter we summarise what capabilities the first and the second phases of the SKA will be able to achieve in its current state of design. We summarise the different cosmological experiments which are outlined in further detail in other chapters of this Science Book. The SKA will be able to produce competitive Baryonic Oscillation (BAOs) measurements in both its phases. The first phase of the SKA will provide similar measurements as optical and IR experiments with completely different systematic effects whereas the second phase being transformational in terms of its statistical power. The SKA will produce very accurate Redshift Space Distortions (RSD) measurements, being superior to other experiments at lower redshifts, due to the large number of galaxies. Cross correlations of the galaxy redshift data from the SKA with radio continuum surveys and optical surveys will provide extremely good calibration of photometric redshifts as well as extremely good bounds on modifications of gravity. Basing on a Principle Component Analysis (PCA) approach, we find that the SKA will be able to provide competitive constraints on dark energy and modified gravity models. Due to the large area covered the SKA it will be a transformational experiment in measuring physics from the largest scales such as non-Gaussian signals from $f_{nl}$. Finally, the SKA might produce the first real time measurement of the redshift drift. The SKA will be a transformational machine for cosmology as it grows from an early Phase 1 to its full power.}

\FullConference{
Advancing Astrophysics with the Square Kilometre Array\\
June 8-13, 2014\\
Giardini Naxos, Italy}

\begin{document}

\section{Introduction}

It has been over a decade since we entered an era of precision cosmology in which precise estimation of the key cosmological parameters is the ultimate goal of cosmological surveys. The advances in our ability to map out the Cosmic Microwave Background (CMB) have lead the way and culminated in the very accurate datasets which are available to us from the Planck mission \cite{2013arXiv1303.5076P}. The CMB is however limited in providing information about the nature of components in the late time Universe given that it is mainly probing the nature of the Universe, some 400,000 years after the Big Bang. The Planck mission is able to provide somewhat better constraints on Dark Energy given that it can also measure the lensing of the CMB, which other earlier missions did not have the capability of doing \cite{2013arXiv1303.5077P}. Even with the lensing of the CMB, Planck is not an ideal mission for obtaining information about the nature of dark energy.

It is with galaxy surveys that the way will be paved in the next decade in terms of improving our current knowledge on the accelerating Universe. Missions that probe both the geometrical nature of the late time Universe as well as the growth of structure will be the missions which, statistically speaking, have the best handle on Dark Energy. Missions such as the SKA and the Euclid Satellite will indeed measure both geometry and growth with galaxy redshift surveys by using techniques such as Baryonic Acoustic Oscillations and Redshift Space Distortion measurements of the Large Scale Structure. There are more novel optimal ways of probing the Large Scale Structure of the Universe with radio telescopes using Intensity Mapping which are summarized in \cite{IMSANTOS}. However in this chapter here we concentrate on the arguably more systematic-free experiment which aims at obtaining full galaxy redshift surveys with the SKA. 

The outline of this summary chapter is as follows. In Section 2 we outline the basic assumptions used for the forecast we have made in the rest of the chapter. In Sections 3 and 4 we summarise the main cosmological probes, namely the Baryonic acoustic oscillations and Redshift space distortion measurements with SKA. In Section 5 we look at the power of an SKA galaxy redshift survey when used in conjunction with other SKA probes to measure Modified Gravity and to calibrate photometric redshifts. In section 6 we look at the science that can be done with an SKA at the largest possible scales by probing non-Gaussianity. In Section 7 we summarise the cosmological redshift drift science that is accessible to a future SKA2. We conclude in Section 8.

\section {The SKA1 and SKA2 galaxy redshift surveys: Number counts and bias}

Neutral hydrogen (HI) is the most abundant element in galaxies, making it a prime candidate to observe them and trace the underlying dark matter. Moreover, with a rest frequency of 1420 MHz, the HI (21 cm) line can be used by radio telescopes to measure the redshift of galaxies to high accuracy and across a large redshift range. The line is however quite weak, requiring highly sensitive telescopes such as the SKA in order to make large sky HI galaxy surveys (so far, only a few galaxies up to $z\sim 0.2$ have been detected \cite{2011ApJ...727...40F}).

Predicting the number of HI galaxies one will be able to observe, as well as the associated bias with respect to the dark matter, are fundamental ingredients in order to forecast the constraining power for cosmology of future radio telescopes. Two components are essential for this prediction: i) the HI content/luminosity of galaxies and ii) the expected sensitivity of the telescope. In the chapter by \cite{santosSKAdndz} we analysed explicitly the sensitivities for different SKA setups (both for phase 1 and the full SKA) and the corresponding number counts and bias. The HI luminosities were obtained using the SAX-sky simulation \citep{2009ApJ...703.1890O}. Translating this into a detection is not completely straightforward since it depends on the source detection algorithm. We assumed that at least two points across the HI line should be measured, in order to have some handle on the typical ``two-horn'' structure of the HI line and the combined measurement of all the points across the line should result in a signal to noise of $10\sigma$ or at least $5\sigma$. To calculate the bias, galaxies from the SAX simulation for a given flux sensitivity and redshift were put in the box for which the power spectrum of the number counts was calculated and compared to the dark matter one. Due to the size limits of the simulation, shot noise can contaminate the measurement so that values at flux sensitivities above $20 \mu$Jy or at very high-z, should be only indicative (this is however not a concern since the shot noise tends to dominate the measurements). 

Given the required balance for cosmological applications between surveying a large sky area to beat cosmic variance and having high galaxy number densities to beat shot noise, we found that the optimal sky area for a SKA1 survey is around 5,000 deg$^2$, while SKA2 should probe the full available sky, $\sim$ 30,000 deg$^2$. Results show that we should find about $5\times 10^6$ HI galaxies with SKA1 (for either MID or SUR) up to $z\sim 0.5$ and about $9\times 10^8$ galaxies with the full SKA up to $z\sim 2$. Obviously results might change a little as the SKA specifications are tuned and different HI simulations are used. These numbers assumed a $10\sigma$ detection with the full SKA while for SKA1 we took a more relaxed requirement of $5\sigma$ in order to allow for a higher detection of galaxies. 
The bias is shown to have a value around 1 at $z\sim 0.7$, increasing with both redshift and assumed flux cut since we will be probing higher mass galaxies.
The numbers required for the forecasting done throughout the chapters can be nicely summarised by the following equations:
\begin{eqnarray}
dn/dz &=& 10^{c_1} z^{c_2} {\rm exp}\left( - c_3 z\right) \label{equ: dndz} \\
b(z) &=& c_4 \exp({c_5z}), \label{bias}
\end{eqnarray}
where $dn/dz$ is the number of galaxies per square degree and per unit redshift. The $c_i$ parameters for each survey should be taken from Table 4 in \cite{santosSKAdndz}. The main conclusion is that although SKA1 will already detect a large number of galaxies if we compare to optical surveys, we can only use it for cosmological applications up to $z\sim 0.4$ due to the sharp decline of the HI galaxy density with redshift. This means that frequencies $\gtrsim 1$ GHz should be enough (i.e. Band 2). On the other hand, SKA2 will push this up to $z\sim 1.7$, requiring a significantly wider band down to $\sim 500$ MHz (or even higher redshifts with Band 1, down to 350 MHz). It will also be able to cover the full visible sky ($\sim$30,000 sq. deg.), making it a prime instrument for cosmological applications.

\section{BAO and distance constraints}

\begin{figure}[t]
\centering{
\includegraphics[width=0.44\columnwidth]{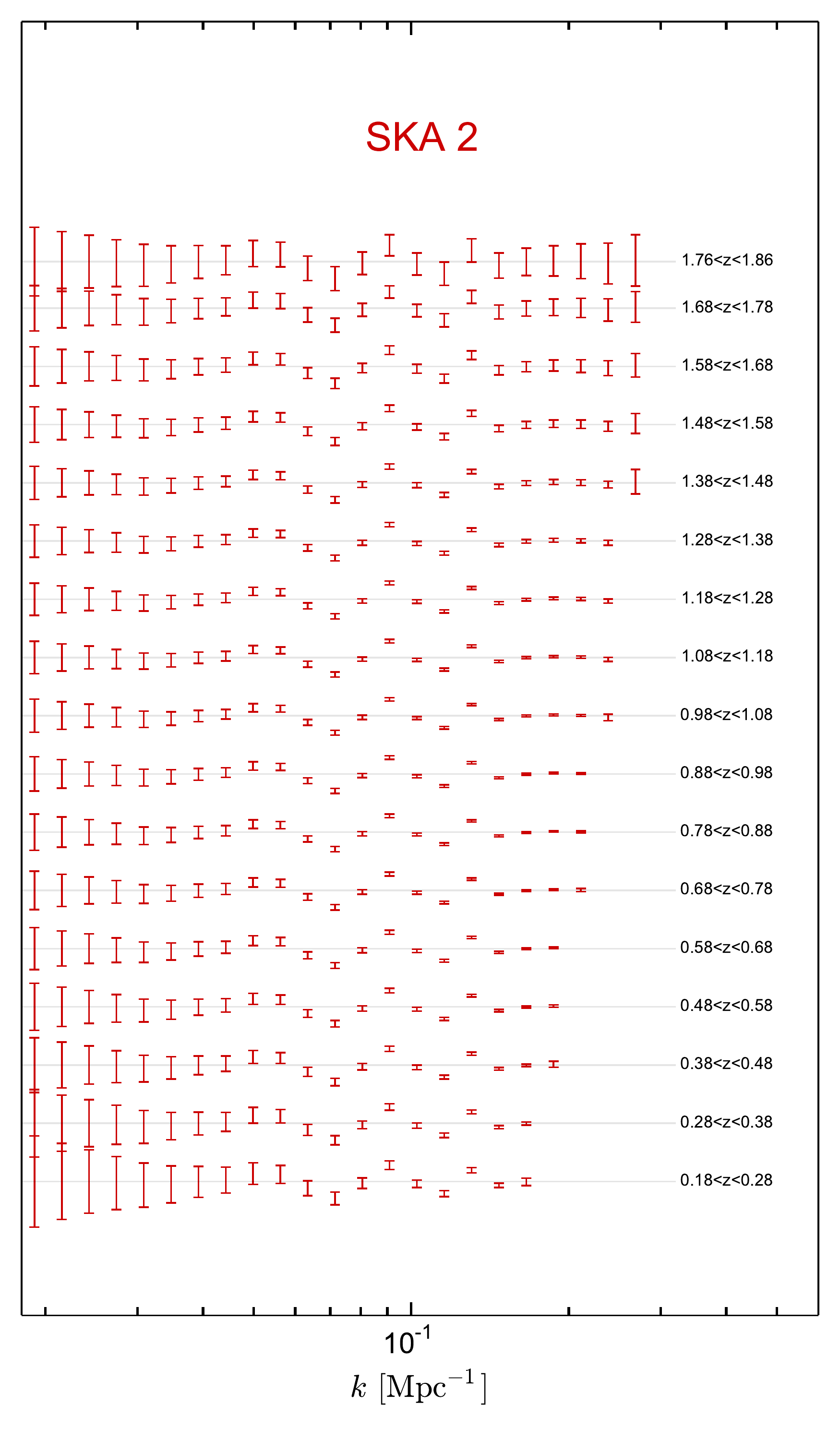}
\includegraphics[width=0.44\columnwidth]{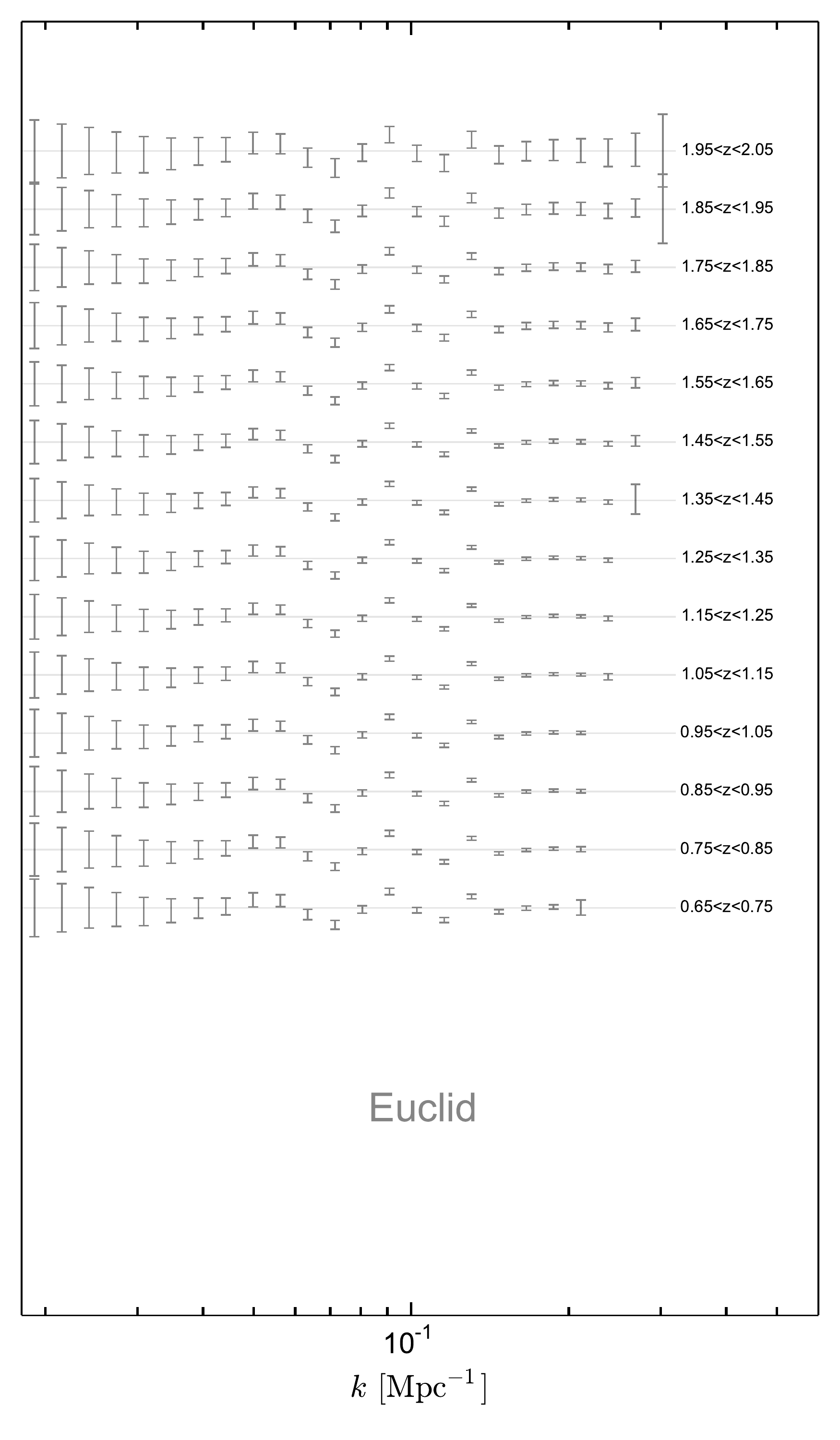}
\caption{Constraints on the baryon acoustic oscillation feature in the power spectrum, as a function of scale and redshift, for SKA2 (left) and Euclid (right). Both the SKA and Euclid will be extremely competitive for such measurements, but the SKA2 survey will have a wider redshift range -- extending to lower redshifts than Euclid -- and will cover a larger survey area.}\label{fig:BAO} }
\end{figure}

\begin{figure}[t]
\centering{
\includegraphics[width=0.85\columnwidth]{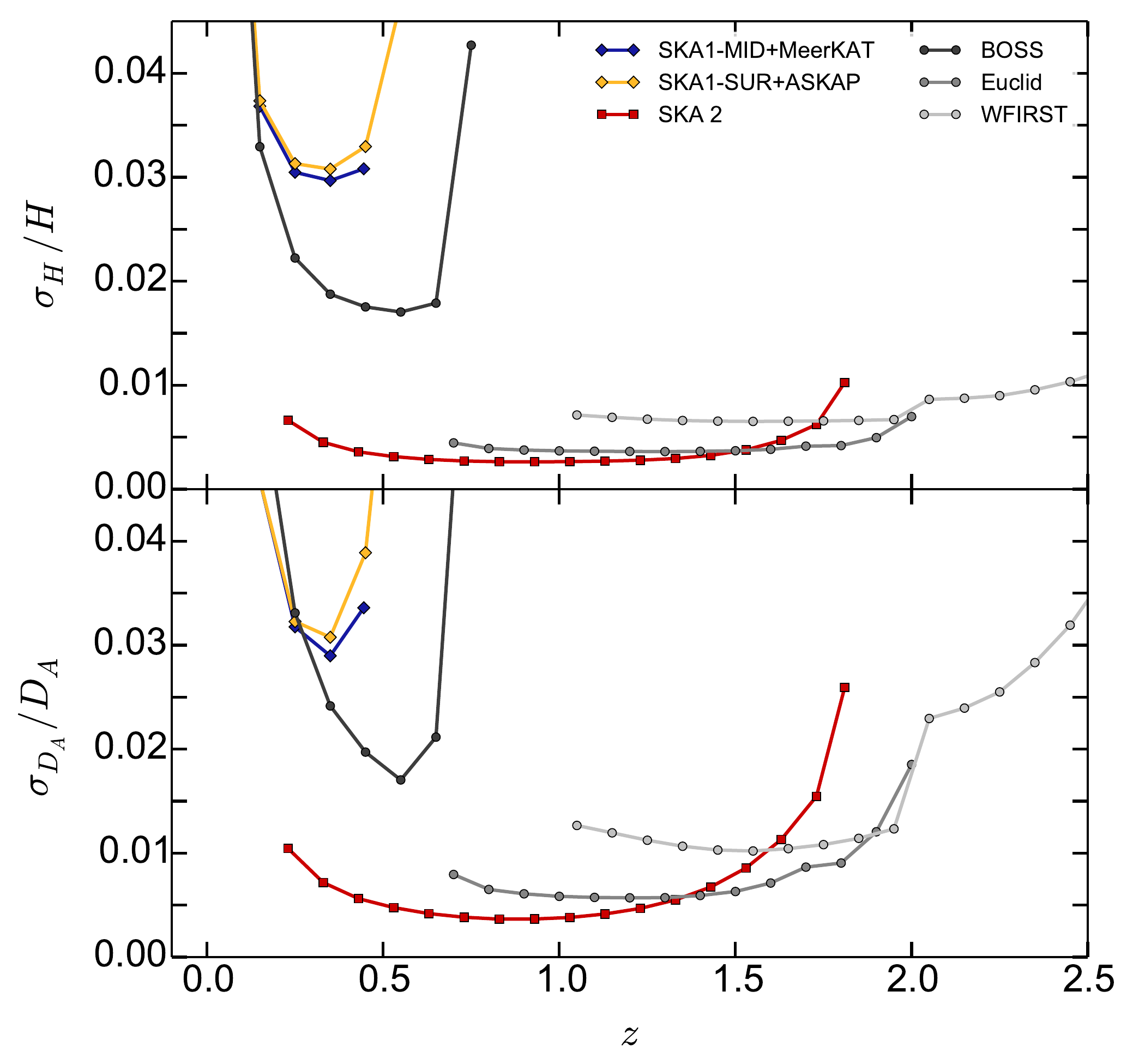}
\caption{Forecast constraints on the angular diameter distance and expansion rate measured from the BAO, RSDs, and broadband shape of the power spectrum. While an SKA Phase 1 galaxy survey will not be competitive, Phase 2 will produce the best constraints of any planned survey on both quantities, up to $z \simeq 1.4$.}\label{fig:hda} }
\end{figure}

Baryon acoustic oscillations (BAO) are the imprints of sound waves -- formed when baryons and photons were coupled at early times -- on the large-scale distribution of matter in the Universe. Because they constitute a preferred scale in the clustering distribution of galaxies, one can use the BAO as a statistical standard ruler, allowing accurate measurements of distances and the expansion rate as a function of redshift, which in turn provide key information on the energy content of the Universe, and the nature of dark energy.

HI galaxy redshift surveys with the SKA will be capable of measuring the BAO in both the radial and tangential directions over an unprecedentedly large volume, taking in around three quarters of the sky out to high redshifts. For Phase 1, a survey with flux sensitivity $\sim 100 \mu\mathrm{Jy}$ over 5,000 sq. deg. will reach $z \approx 0.5$, producing broadly comparable constraints to BOSS \cite{2012MNRAS.427.3435A}. As several large planned galaxy surveys (e.g. Euclid) obtain spectra in the infrared part of the spectrum, they are unable to obtain significant numbers of galaxies at very low redshift. As such, a HI survey will be highly complementary, providing a useful low-redshift anchor to the Euclid galaxy redshift survey (which will cover 15,000 sq. deg. from $0.7 \lesssim z \lesssim 2$), for example (Fig. \ref{fig:BAO} and Fig. \ref{fig:hda}).

\begin{figure}[t]
\centering{
\includegraphics[width=0.90\columnwidth]{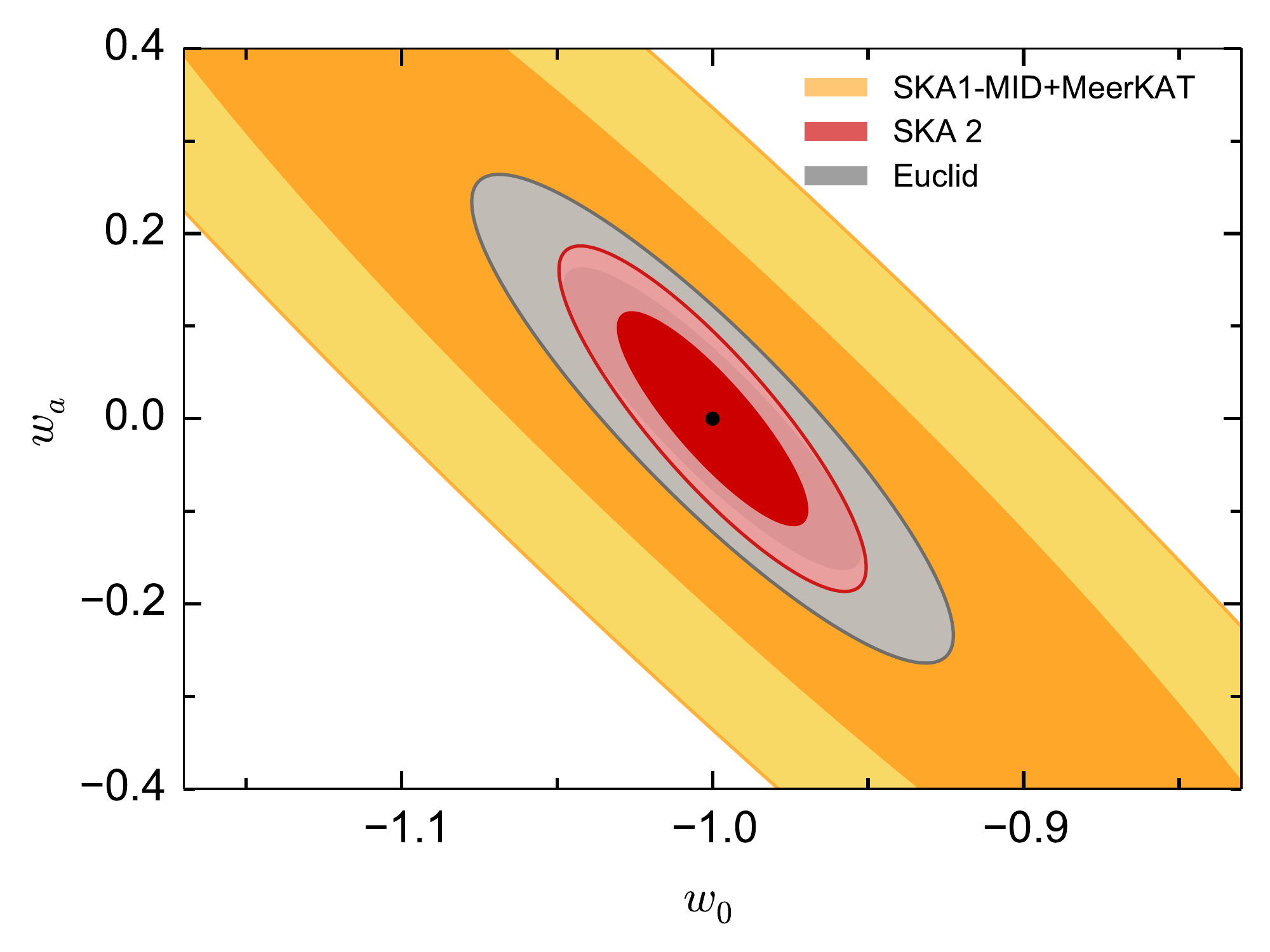}
\caption{Constraints on the dark energy equation of state in the $(w_0, w_a)$ parametrisation, using BAO measurements as well as Planck CMB and BOSS low-redshift BAO priors. A ``billion galaxy'' redshift survey with SKA2 will be significantly more powerful than the Euclid galaxy survey, mostly due to its extended redshift coverage and larger survey volume. Surveys with SKA1 will be complementary to optical and IR surveys, as they will cover a large area of the sky with high density at low redshift, but cannot compete in terms of raw dark energy constraints as they do not extend to high enough $z$.}\label{fig:w0wa} }
\end{figure}

A Phase 2 survey down to a threshold of $\sim 5 \mu\mathrm{Jy}$ over the same area will result in a dramatically extended redshift range, providing $\sim 0.3\%$-level constraints on both the angular diameter distance and expansion rate in the range $0.4 \lesssim z \lesssim 1.3$. This is sufficient to surpass the precision of a Euclid galaxy redshift survey. The galaxy selection function drops out at higher redshifts, though, leaving intensity mapping as the SKA's principal BAO probe at $z \gtrsim 1.5$. Further details on BAO measurements, including potential systematic and results from existing surveys, can be found in \cite{SKABAO}. We can see the full forecasts on the equation of state for dark energy in Fig.\ref{fig:w0wa} where the statistical power of SKA1 and SKA2 is compared to that of the Euclid mission in constraining dark energy. All the above assumes 10,000 hours of survey time; unlike Euclid, however, the SKA is a permanent facility that will be available for the best part of the next century, making it possible (in principle) to considerably improve on the above forecasts with longer surveys.

\section {RSD constraints}

Baryonic Acoustic Oscillations can only provide information which relates to the geometry of the late time Universe. There are however several competing theories of Dark Energy which are degenerate in their prediction of the late time behavior of the accelerated expansion. It is possible however to distinguish such theories by looking at how structure grows in such models. The peculiar motion of galaxies can be an extremely good probe of gravitational collapse and hence growth. Hence if the dynamics of galaxy motions can be measured by looking at the statistical distribution of galaxies we can use such data to further constrain models of dark energy, by looking at modifications to the full shape of the power spectrum (or correlation function) of galaxies. This is usually expressed in terms of measurements of the linear growth rate, $f$, which denotes the logarithmic derivative of the matter density contrast with respect to the scale factor. Although there are complications with RSD measurements which include knowledge of the galaxy bias, and in some cases non-linear and large-scale modeling, RSDs remain one of the best ways of producing measurements of modified theories of gravity. Given that redshift space distortions manifest themselves in anisotropies in the measured power spectrum signal as a function of the angle we have with the line of sight, other effects such as the Alcock-Paczynski (AP) effect weakens the constraints that one may get from Redshift space distortions. However, by measuring the magnitude of this effect we may produce further geometrical measurements of the AP factor, $F$, which is related to a ratio of the angular diameter distance and the Hubble factor.

\begin{figure}[t]
\centering{
\includegraphics[width=0.8\columnwidth]{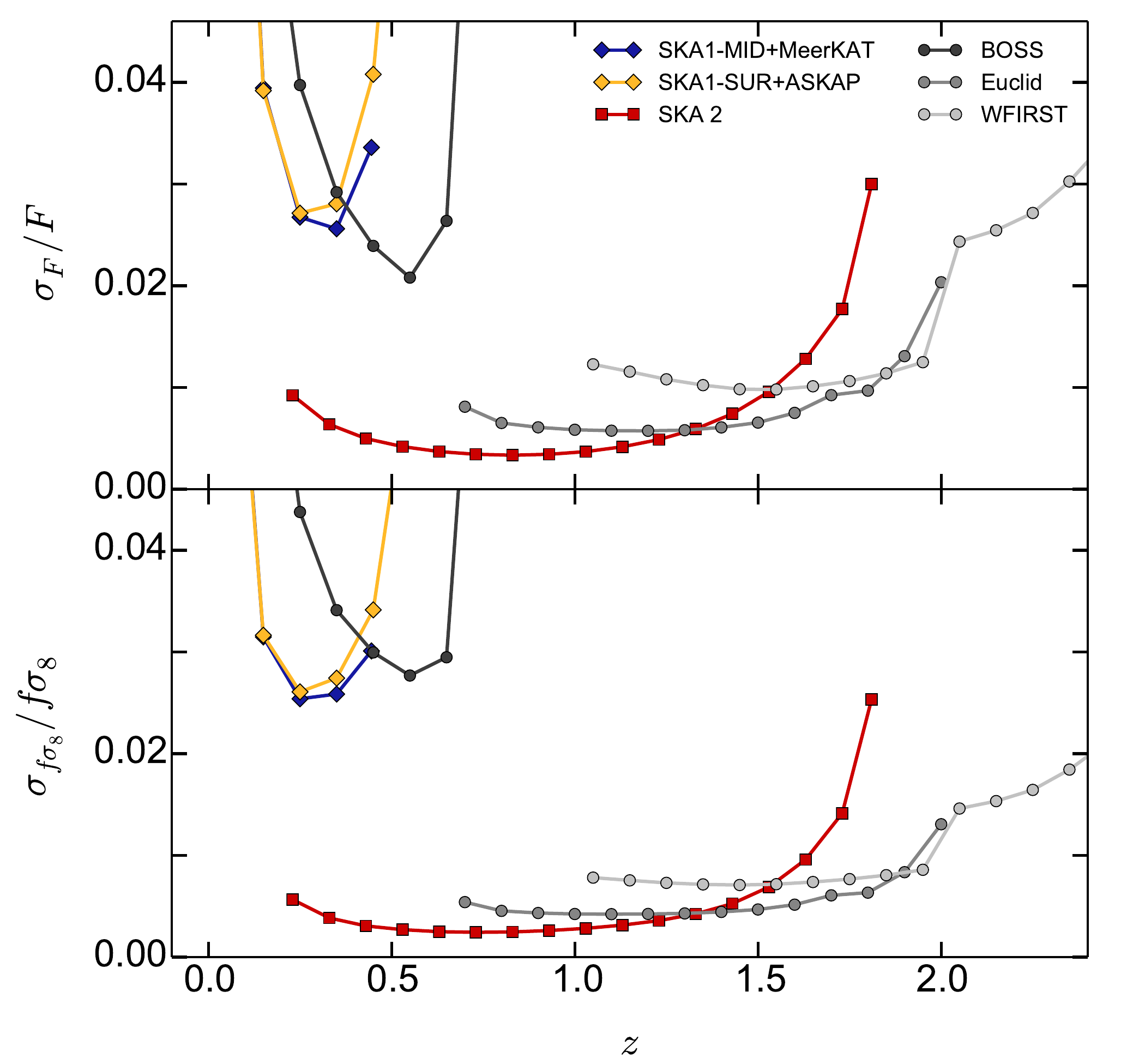}
\caption{Sensitivity to redshift space distortions of HI galaxy surveys with Phase 1 and 2 of the SKA, compared with other galaxy surveys. The upper panel shows the Alcock-Paczynski parameter, $F(z) = (1+z)D_A(z)H(z)/c$, and the lower panel shows the growth rate observable $f\sigma_8$, which is the product of the linear growth rate, $f(z)$, and the normalisation of the power spectrum, $\sigma_8(z)$.}\label{fig:rsd} }
\end{figure}

\begin{figure}
\includegraphics[width=1.0\columnwidth]{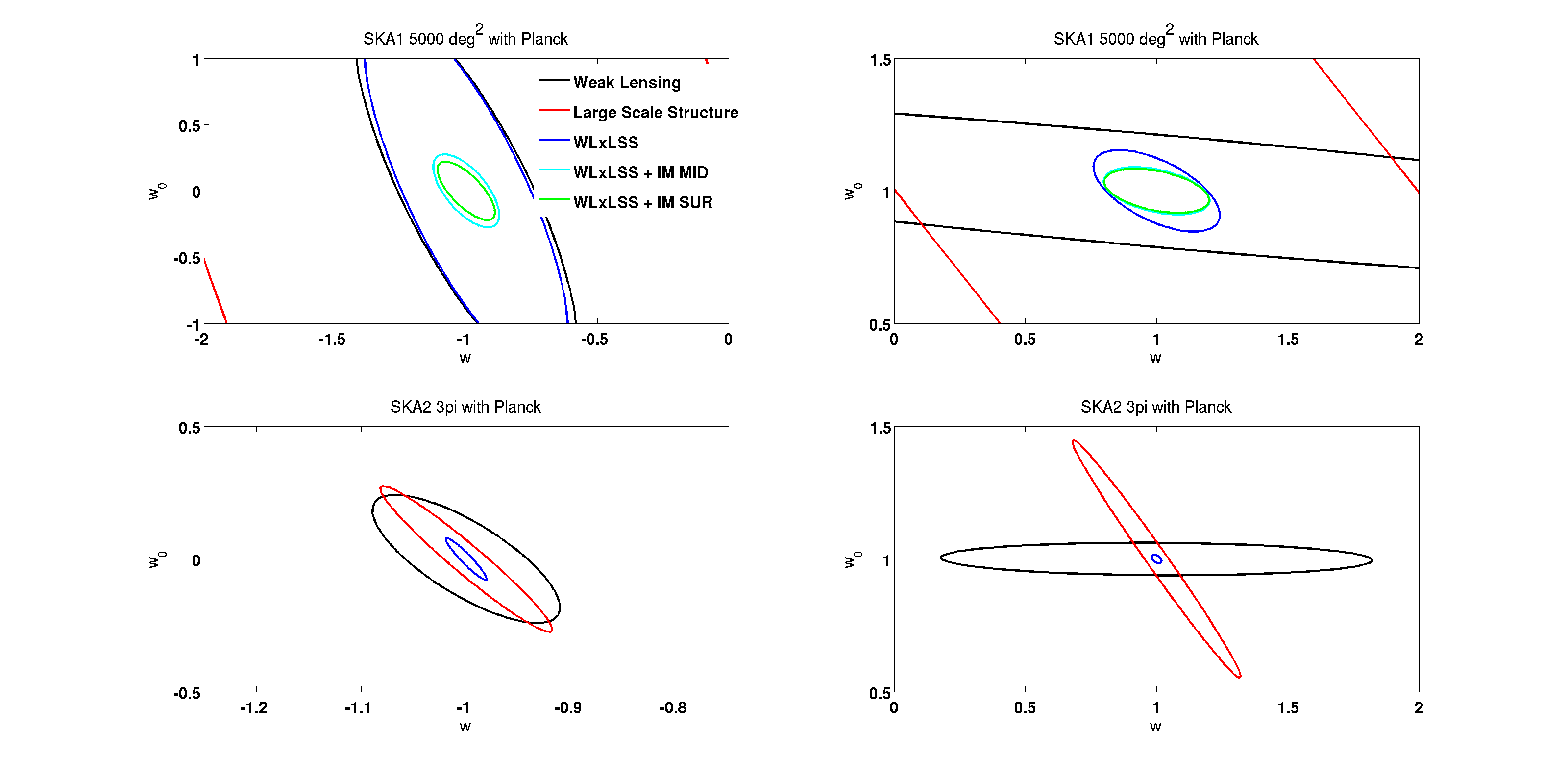}
\caption{Improvement on the Figures of merit of dark energy and Modified gravity using the cross correlations of the spectroscopic SKA sample with other SKA surveys. Constraints on dark energy [left panels] and deviations from GR [right panels] for SKA1 over 5,000deg$^2$ [top panels] and SKA2 over 30,000deg$^2$ [bottom panels] including Planck priors. Black ellipses show photometric WL constraints only. Red ellipses show spectroscopic LSS constraints only. Blue ellipses show the combination of WL and LSS including cross-correlations. Cyan ellipses show this WLxLSS constraint combined independently with an SKA intensity mapping (IM) survey using the MID instrument and the green contours the same but with the SUR instrument. All constraints are 68\% confidence contours.} \label{fig:DE_MG_crosscorr}
\end{figure}

The vast numbers of galaxies found in the SKA Phase 1 and 2 surveys will allow for a very accurate measurement of the redshift space distortion factor, $f$,  as well as the AP distortion factor, $F$. As we can see from Fig.\ref{fig:rsd}, the SKA1 survey is competitive at low redshift, and can provide complementary constraints to the Euclid survey over a very large area in the sky. This is due to the fact that the Euclid mission is targeting H$\alpha$ lines to secure redshifts and doing so at higher redshifts. We can also see that as we move on to a survey with SKA3, the constraints from the SKA are better than Euclid-type experiments for redshifts $z \lesssim 1.2$. A survey with SKA2 would be able to measure redshift space distortions to better than $0.5\%$ precision in the range $0.4<z<1.3$. This ``billion galaxy'' survey can also measure the AP factor to a similar level over the same redshift range. An SKA1 survey would be able to measure both distortion parameters to a little over 2\% accuracy out to $z \simeq 0.5$. Further details on RSD measurements from the SKA, including potential systematic and results from existing surveys, can be found in \cite{SKARSD}.

\section {Cross correlation science with the SKA}

SKA will be able to deliver multiple surveys of the same large-scale structure (LSS) fields which makes it a perfect opportunity to benefit from the power of cross-correlating different cosmic probes. The combination of Intensity Mapping (IM), Weak Gravitational Lensing (WGL) of distant galaxies from the continuum survey and a HI galaxy survey which delivers precise redshift estimates makes SKA an incredibly exciting project for exploring cross-correlation science.

The combination of an SKA galaxy redshift survey with spectroscopic quality redshifts and an SKA WGL survey with photometric redshifts obtained from some external source is particularly powerful. The LSS survey has extremely high redshift resolution but uses galaxies which are biased tracers of the underlying dark matter distribution. The WGL survey has poor redshift discrimination but is unbiased and has a higher number density of sources than the galaxy survey. Together they comprise a joint data set with excellent redshift information and good control of galaxy bias.

The power of cross-correlation comes from the fact that the probes are both sensitive to the same underlying density field. The cross-correlation signal is immune to certain systematic effects that limit each autocorrelation, allowing them to be controlled. Each cosmological probe is sensitive to particular combinations of cosmological parameters. By accessing the underlying field in different ways the cross-correlation can help break degeneracies between those parameters, producing markedly tighter constraints.

In Fig. \ref{fig:DE_MG_crosscorr} we show an example of the powerful constraints that can be obtained from the cross-correlation of SKA LSS, WGL and IM surveys. We show constraints on dark energy in the left panels and deviations from GR in the right panels. The top panels show constraints from an SKA1-type survey and the bottom panels from a full SKA survey.

We show constraints from the photometric WGL and spectroscopic LSS surveys on their own. We also show the combination of WGLxLSS including all cross-correlations and include IM priors for the SKA1 survey. The joint constraints on DE are several times tighter than any single probe. All constraints include priors from the Planck CMB temperature autocorrelation.

The constraints on gravity are particularly impressive, with the cross-correlation yielding measurements that are orders of magnitude more accurate than either probe alone. This is due to the combination of one probe that uses non-relativistic tracers (i.e. galaxies in the LSS survey) and one that uses relativistic tracers (WGL measures distortions due to the deflection of light). Relativistic tracers are sensitive to the sum of the metric potential, $\Phi + \Psi$, while non-relativistic tracers respond to the Newtonian potential, $\Psi$, alone. In GR these potentials are equal but Modified Gravity (MG) models generically produce different metric potentials. The combination of both types of probe breaks strong degeneracies in MG theories, producing dramatically improved sensitivity. Further details on the improvement are presented in detail in \cite{SKACROSS}.

\section{Model-independent constraints on Dark Energy and Modified Gravity}

As a powerful redshift survey, the SKA HI galaxy survey will provide competitive constraints on the equation-of-state $w(z)$ of dark energy and the $\mu(k,z),\gamma(k,z)$ functions, which quantify the effect of the modification of gravity. In the $\Lambda$CDM model, $w=-1$ and $\mu=\gamma=1$. Any departure of these functions from these fiducial values would imply the dynamics of dark energy, or the breakdown of general relativity (GR), or both. 

\begin{figure}[t]
\centering{
\includegraphics[width=0.95\columnwidth]{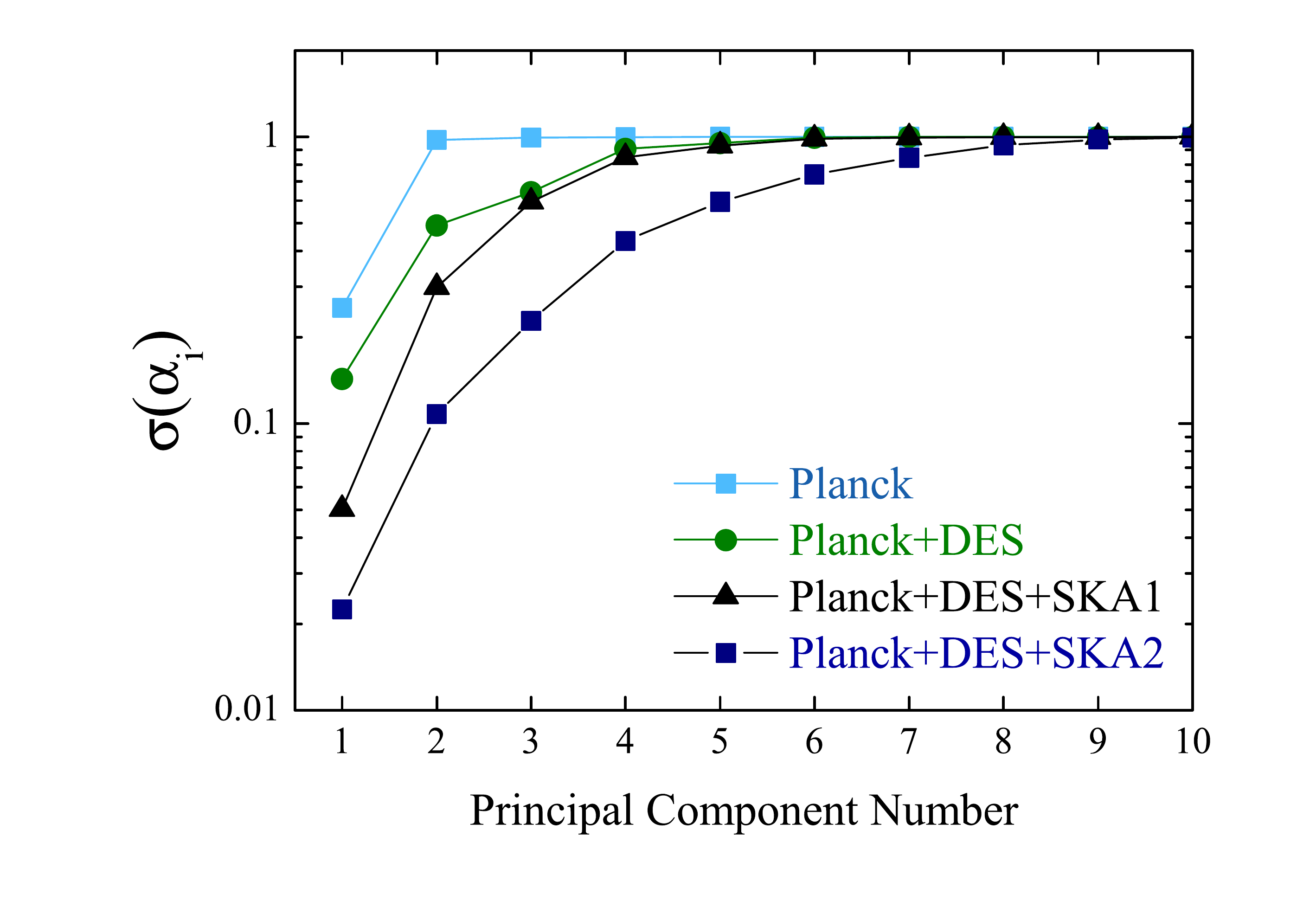}
\caption{The forecasted 68\% CL measurement error on $\alpha_i$, the coefficient of the $i$th principal components of $w(z)+1$, namely, $w(z)+1=\sum_i \alpha_i e_i(z)$, using different data combinations illustrated in the legend. A weak prior of $\sigma(w(z))<1$ was assumed.}\label{fig:wPCA} }
\end{figure}

\begin{figure}[t]
\centering{
\includegraphics[width=0.95\columnwidth]{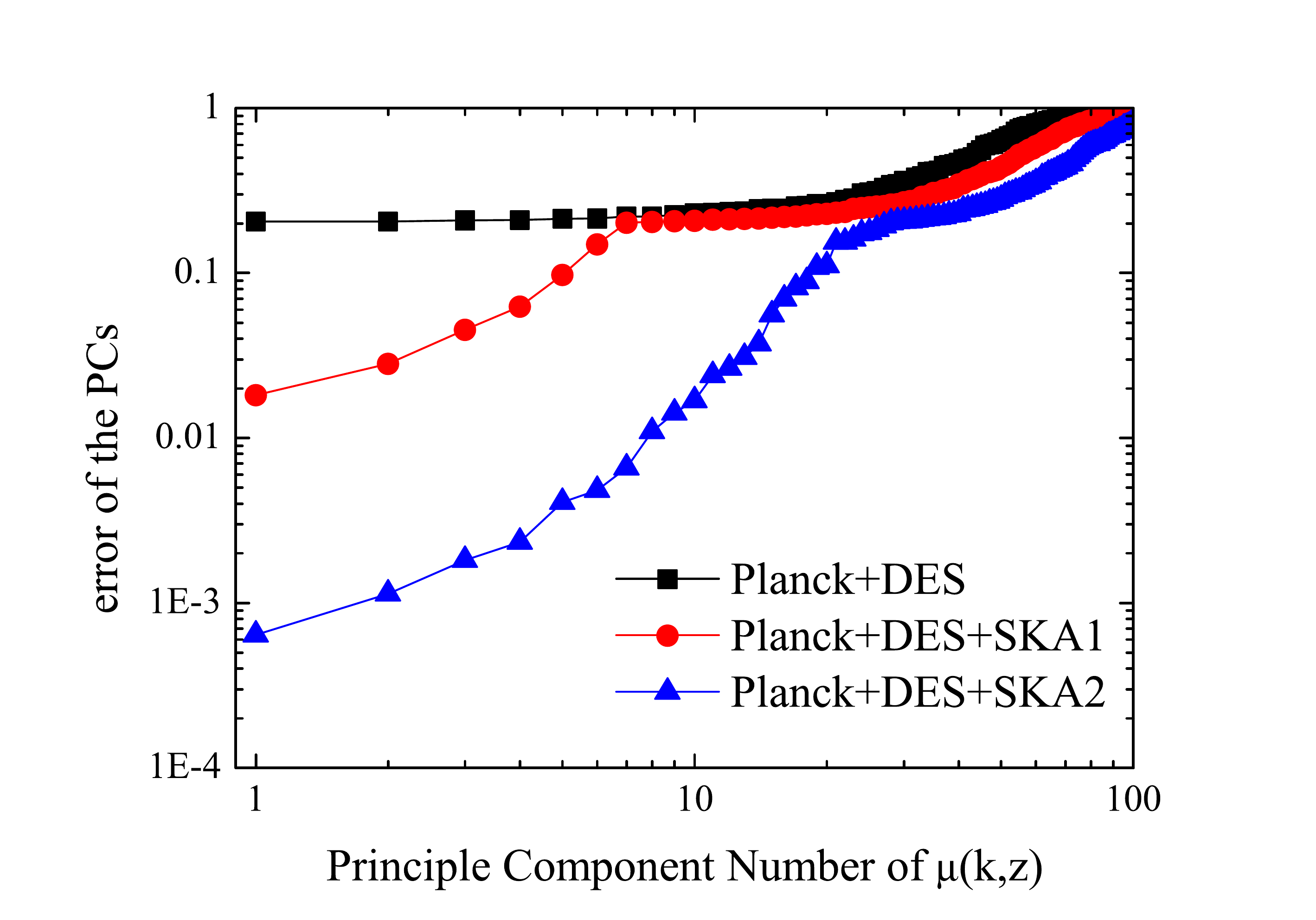}
\caption{The forecasted 68\% CL error on the coefficients of the principal components of $\mu(k,z)$ for different data combinations shown in the legend. The function $\mu$ quantifies the effect of modified gravity on structure formation on linear scales. Specifically, the Poisson equation reads $k^2\Psi=4\pi{G}\mu(k,z)\delta\rho$. }\label{fig:MGPCA} }
\end{figure}

The easiest way to study the observational constraints on the $w,\mu, \gamma$ functions is to parameterise them first, and to constrain these parameters using observations. However, this approach might suffer from the {\it theoretical bias}, {\it i.e.}, the result can be biased if the {\it ad hoc} parametrisation is inappropriate. Nonparametric methods, {\it e.g.}, the Principle Component Analysis (PCA) approach, can not only minimise the theoretical bias, but also provide invaluable information on the detectability of dark energy dynamics and the modification of gravity. For a detailed study of the PCA forecast of dark energy and modified gravity using the SKA HI survey, we refer the readers to the SKA PCA chapter \cite{SKAPCA}. Here we only summerise the main result in Figs \ref{fig:wPCA} and \ref{fig:MGPCA}. 

Fig \ref{fig:wPCA} shows the error of the eigenmodes of $w(z)$ using different data combinations. Combining with the simulated Planck and DES data, we find that SKA Phase 1 (SKA1) and SKA Phase 2 (SKA2) can well constrain $3$ and $5$ eigenmodes of $w(z)$ respectively. The errors of the best measured modes can be reduced to $0.04$ and $0.023$ for SKA1 and SKA2 respectively, making it possible to probe for dark energy dynamics.  

Fig \ref{fig:MGPCA} is a similar plot for the $\mu(k,z)$ function, which quantifies the effect of modified gravity. As shown, SKA1 and SKA2 can constrain $7$ and $20$ eigenmodes of $\mu(k,z)$ respectively within the 10\% sensitivity level. Especially 2 and 7 modes can be constrained within sub percent level using SKA1 and SKA2 respectively. This is a significant improvement compared to the combined datasets without SKA.

\section{Cosmology on the largest scales with the SKA}

All-sky SKA surveys, in HI threshold and intensity mapping \cite{IMSANTOS} and in continuum \citep{CONTJARVIS} , will probe the biggest volumes ever of large-scale structure in the Universe. This will allow a major advance in tackling two of the main questions in cosmology: Are the fluctuations generated in the early Universe non-Gaussian? Does General Relativity (GR) hold on the largest scales?

The most stringent constraints currently available on primordial non-Gaussianity come from the Planck CMB experiment \citep{2013arXiv1303.5084P}. Future CMB data on polarization will improve these
results, but probably not significantly. Surveys of the matter distribution are the new frontier for non-Gaussianity (see e.g. \cite{2012MNRAS.422.2854G,2014PhRvD..89b3511G}), and the SKA has the potential to deliver game-changing constraints, given the ultra-large scales that it probes \citep{Camera:2014bwa}. Indeed, non-Gaussian correction to the clustering of biased tracers of the underlying DM distribution occur on extremely large scales \citep{2008PhRvD..77l3514D,2008ApJ...677L..77M}. Since Planck detected no non-Gaussianity larger than $|f_{\rm NL}|\sim10$, we need to probe huge volumes of the cosmic large-scale structure if we want to be able to detect such a tiny deviation from Gaussianity. This is because the larger the surveyed scale, the tighter the constraint on primordial non-Gaussianity \citep{2013PhRvL.111q1302C}. This expectation is supported by recent forecasts of the constraining power of different SKA surveys \citep{Camera:SKACHAP}.
Another advantage for SKA arises from the possibility of discriminating between different source types, allowing the use of the so-called multi-tracer
technique to reduce cosmic variance and thus further increase the constraining capabilities on very large scales \citep{2009PhRvL.102b1302S,2013MNRAS.432..318A,2014MNRAS.442.2511F}.

Tests of GR on cosmological scales can only be based on a combination of observations of the large-scale structure. Current constraints are weak, but with its huge volumes and multiple probes, the SKA should lead the next generation of tests. In addition, we can strengthen the current tests of dark energy and modified gravity models by extending these tests to much larger scales, increasing the statistical power of the observations and improving constraints on any scale dependence of deviations from GR.

Probing ultra-large scales includes a theoretical challenge that has only been recently recognized. On very large scales, relativistic effects on the galaxy over-density  also become important. (This is also true for the HI brightness temperature fluctuations, but the relativistic effects are suppressed in intensity mapping). They arise since we observe on the past lightcone and they can significantly change the predictions of the standard Kaiser approximation. In addition to the standard redshift-space distortions and the weak lensing magnification, there are Doppler, Sachs-Wolfe, integrated SW and time-delay contributions, which can become significant on horizon scales. It is therefore necessary to include these effects in predictions that will be tested against observations. 

The SKA has the potential to detect these relativistic effects, thus providing a clear signal of GR---or an indication of its violation on the largest scales. Furthermore, the amplitude of the relativistic effects can be of the same order as primordial local non-Gaussianity that is consistent with Planck's constraints. 

Primordial non-Gaussianity is a powerful test of inflationary models---and Planck has already ruled out those models that generate large non-Gaussianity. The key target is the simple single-field inflation models. These models generate negligible non-Gaussianity, but a nonlinear GR correction to the Poisson equation means that large-scale structure would measure $f_{\rm NL}\sim -2$.
(We will discuss this and the new techniques to improve constraints on primordial non-Gaussianity via a proper accounting of the relativistic effects in detail in \cite{Camera:SKACHAP}.) SKA HI galaxy redshift surveys will greatly help on this effort. Indeed, \cite{Camera:2014bwa} showed that a HI galaxy redshift survey covering the range $0<z\leq3$ is able, thanks to the huge volume it probes, to put the currently most stringent constraint on $f_{\rm NL}$ from a single threshold tracer, i.e.\ $\sigma(f_{\rm NL})=1.54$. Moreover, their analysis has been performed for the first time by fully accounting for the GR effects described above. More specifically, the numbers used for the bias and galaxy redshift distribution for different flux cuts were taken directly from simulations. The other parameters required for the relativistic corrections were in turn directly derived from these parameters, which allowed for a fully consistent analysis of the fluctuations on ultra-large scales taking into account the GR corrections.

\section{Time domain cosmology with SKA}

Real time cosmology will be possible with the SKA due to the superior
survey capabilities and the ability to measure Milk-Way type galaxies
up to redshifts of 1. The basic experiment is to detect the change in
redshift of individual galaxies caused by the expansion of the
Universe with two epochs of observations. Various
observables can be used to describe the expansion history of the
Universe (see e.g. \cite{2002ApJ...565....1G}), but apart from
the redshift most of these observables are out of reach by the current
technical capabilities of the SKA. In order to measure the redshift
drift caused by the acceleration of the ``near-by''  Universe the
frequency resolution of the SKA needs to be adapted, which is a
feasible task in synthesis interferometer such as the SKA and could be
realised via a software solution.

\begin{figure}
  \centering
  \includegraphics[width=13cm]{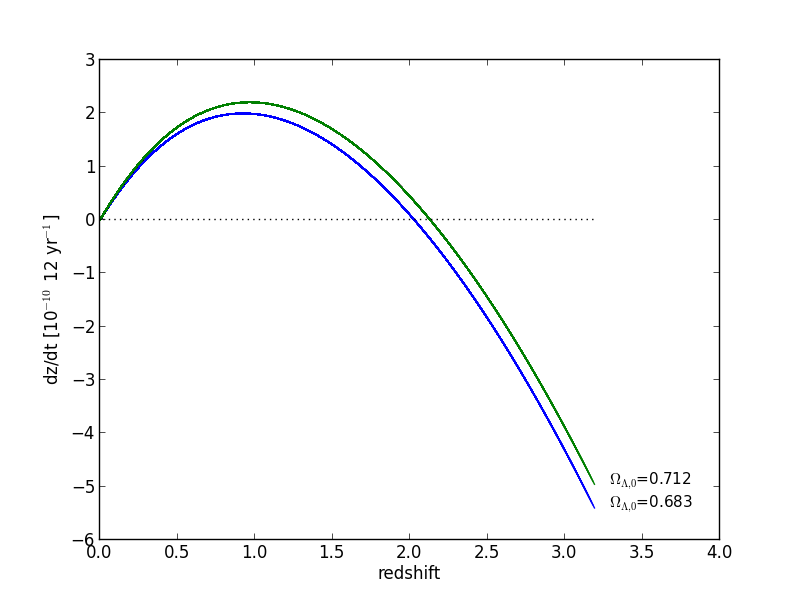}

  \caption{The expected redshift drift for various $\Lambda$CDM
    cosmologies ($\Omega_{\Lambda}$\,=\,0.712 WMAP,
    $\Omega_{\Lambda}$\,=\,0.683 Planck) assuming two epochs of
    observations within 12\,years.  }
  \label{FIG:dzdt}
\end{figure}

The idea to measure the expansion rate of the Universe via redshifts
has been explored in 1962 by Sandage. At that time the
technological limitations made these measurements out of reach. It
took more than 30 years until the idea was revisited by Loeb in
1998, who proposed to use the Lyman-alpha forest absorption lines
toward quasars to measure the expansion rate of the Universe. This
paper was the basis to investigate these kind of measurements by
optical facilities like the E-ELT and experiments like this are
nowadays referred to as CODEX-like experiments (see e.g. 
\cite{2005Msngr.122...10P}). Unfortunately this test is greatly affected by the
limitations introduced by the atmosphere, which is opaque for
Lyman-alpha photons from redshifts z\,$\geq$\,1.7. In comparison the
redshift estimates performed by the SKA will not have such limitations
and it has been realised that a statistical detection of the cosmic
acceleration would be possible with the SKA at redshifts below a redshift of one
using Milky-Way type galaxies \citep{HANSPROC}. Based on the
cosmological parameter measured by the WMAP and Planck (\cite{2013ApJS..208...19H}, \cite{2013arXiv1303.5077P}) a maximum redshift drift of about
$2\cdot 10^{-10}$ can be expected after an observing period of
12~years (see Fig. \ref{FIG:dzdt}). Furthermore, based on the current
$\Lambda$CDM model of the Universe the contribution of the cosmic
acceleration will change with redshift and the ultimate experiment to
test different cosmological models would be to trace the redshift
drifts at various redshifts. Investigating the sensitivity estimates
and the number counts of the expected HI galaxies, it could be shown
that the number counts are sufficiently large to compensate for the
uncertainties of the measurements and therefore permit a statistical
detection of the redshift drift per redshift bin.  These
investigations also indicated that this experiment does not necessarily
require the full SKA sensitivity, it could be performed already by 2
shallow HI $3/4$-sky surveys built up by 20~sqdeg pointings observed
for 1 hour.  Due to the relatively short survey duration of about a
year per epoch and the anticipated life time of the SKA of about
50~years, the SKA offers the unique opportunity to perform this
experiment five or even more times. Such experiments will open up a
new path in precision cosmology enabling the direct and model
independent measure of the jerk term of the acceleration.  In
addition, the combination of the measurements of the redshift drift in
redshift space by the SKA and the E-ELT, both are sampling different
redshift ranges, are the only experiments that will fully trace the
evolution of the dark matter in the Universe in a completely model
independent way.

\section{Conclusions}

We have summarized the cosmological capabilities for the SKA to perform galaxy redshift surveys. The SKA phase one will be able to map millions of galaxies up to redshift of the order of $z\sim 0.5$ whereas the full SKA, will allow us to have a billion galaxy redshift survey over most of the sky. We have summarized the forecasts for BAO and RSD measurements and shown that the SKA1 is complementary to other state of the art facilities, while the full SKA provides transformational science and greater accuracy than other space missions in the next decade. We have outlined how well the SKA can use its galaxy redshift surveys to provide Modified Gravity constraints as well as calibrating photometric redshifts using cross correlations with other SKA and optical samples. We have summarized how well SKA can do to constrain the dynamics of dark energy and modified gravity using the nonparametric method (the PCA approach). A full SKA survey will be able to yield cosmology at the largest scales providing measurements of $f_{nl}$ to better accuracy than the Planck satellite, with different physical probes. We have also summarized the capabilities of the SKA over a ten year period to map out the redshift drift in real time. In summary the SKA will provide transformational science in cosmology.

\bibliography{references}





\end{document}